\documentclass[12pt]{article}
\usepackage[margin=2.5cm]{geometry}
\usepackage{amsmath,amssymb,extarrows,mathtools,graphicx,subfigure,setspace,bbold}
\usepackage{xcolor}
\usepackage{dsfont}
\usepackage{tikz}
\usetikzlibrary{decorations.pathmorphing}
\usetikzlibrary{decorations.shapes}
\usetikzlibrary{arrows,decorations.pathreplacing}
\usepackage{hyperref}
\usepackage{mathrsfs}
\usepackage{hyperref}
\hypersetup{colorlinks=true, linkcolor=blue, citecolor=magenta, linktoc=page}
\usepackage[square,numbers,sort,compress]{natbib} 
\numberwithin{equation}{section} 

\usepackage{soul} 
\newcommand{\be}{\begin{equation}}
\newcommand{\ee}{\end{equation}}
\newcommand{\bea}{\begin{eqnarray}}
\newcommand{\eea}{\end{eqnarray}}
\newcommand{\bse}{\begin{subequations}}
	\newcommand{\ese}{\end{subequations}}
\newcommand{\beqa}{\begin{eqnarray}}
\newcommand{\eeqa}{\end{eqnarray}}
\newcommand{\beqar}{\begin{eqnarray*}}
	\newcommand{\eeqar}{\end{eqnarray*}}
\newcommand{\bi}{\begin{itemize}}
	\newcommand{\ei}{\end{itemize}}
\newcommand{\bn}{\begin{enumerate}}
	\newcommand{\en}{\end{enumerate}}

\newcommand{\ba}{\begin{array}}
	\newcommand{\ea}{\end{array}}
\newcommand{\bc}{\begin{center}}
	\newcommand{\ec}{\end{center}}
\newcommand{\nnr}{\nonumber \\}
\newcommand{\nn}{\nonumber}

\newcommand{\ie}{{\em i.e.}\ }
\newcommand{\eg}{{\em e.g.} }

\def\l3{\ell_3}

\def\a{\alpha}
\def\a1{a_0}
\def\rh{\rho_0}
\def\m{\tilde{m}}

\definecolor{darkgreen}{rgb}{0,0.3,0}
\definecolor{mgreen}{rgb}{0,0.6,0}
\definecolor{darkblue}{rgb}{0,0,0.7}
\definecolor{darkred}{rgb}{0.7,0,0}

\begin{document}
\setstcolor{red} 

	\begin{titlepage}
		
		\begin{flushright}\vspace{-3cm}
			{\small
				IPM/P-2017/011 \\
				\today }\end{flushright}
		\vspace{0.5cm}

		\begin{center}
			\centerline{{\fontsize{18pt}{12pt}\selectfont{\bf{AdS$_{3}$ to dS$_{3}$ transition in the near horizon }}}} \vspace{8mm}
			\centerline{{\fontsize{18pt}{12pt}\selectfont{\bf{of asymptotically de Sitter solutions}}}} \vspace{10mm}
\vspace{6mm}

\large{\bf{S. Sadeghian\footnote{e-mail: ssadeghian@ipm.ir} and  M.H. Vahidinia\footnote{e-mail:
vahidinia@ipm.ir}  }}
\\

\vspace{5mm}
\normalsize
\bigskip\medskip
{ \it School of Physics, Institute for Research in Fundamental
Sciences (IPM),\\ P.O.Box 19395-5531, Tehran, Iran}\\
			
		\end{center}
		\begin{abstract}
			\noindent

We consider two solutions of Einstein-$\Lambda$ theory which admit the extremal vanishing horizon (EVH) limit, odd-dimensional multi-spinning Kerr black hole (in the presence of cosmological constant) and cosmological soliton. We show that the near horizon EVH geometry of Kerr has a 3d maximally symmetric subspace whose curvature depends on rotational parameters and the cosmological constant. In the Kerr-dS case, this subspace interpolates between AdS$_3$, 3d flat and dS$_3$ by varying rotational parameters, while, the near horizon of the EVH cosmological soliton always has a dS$_3$. The feature of the EVH cosmological soliton is that it is regular everywhere on the horizon. In the near EVH case, these 3d parts turn into the corresponding locally maximally symmetric spacetimes with a horizon: Kerr-dS$_3$, flat space cosmology or BTZ black hole. We show that their thermodynamics match with the thermodynamics of the original near EVH black holes. We also briefly discuss the holographic 2d CFT dual to the near horizon of EVH solutions.
		\end{abstract}
		
		\vspace{0.5in}

	\end{titlepage}
	\setcounter{footnote}{0}
	\renewcommand{\baselinestretch}{1.05}  
	
	\addtocontents{toc}{\protect\setcounter{tocdepth}{2}}
	\tableofcontents
	

	
\section{Introduction}
Exploring the classical and semi-classical aspects of solutions to gravitational theories may provide a good framework to better understand the nature of gravitational fields and the origin of spacetime, such as the casual structure of spacetime in relativistic theories.
In this way, studying the stationary black hole (brane) solutions with their non-trivial causal structures is distinguished. It is known that at the classical level they perform some thermodynamic-like behavior which can be promoted to the real thermodynamics at the semi-classical regime.

Studying black hole physics involves finding and classifying black hole solutions to different gravitational theories in diverse dimensions. Various limits of these black hole solutions and their thermodynamic properties are interesting. In particular,
studying the near horizon  limit of extremal black branes and black holes (which have vanishing surface gravity) have shed light on black hole physics and  (quantum) gravity. The main property of this limit is based on the symmetry enhancement of near horizon geometries at the extremality. The most well-known example is AdS/CFT where the emergence of an AdS throat in the near horizon of  extremal p-branes  inspires a holographic duality between gravity on AdS$_{p+2}$ space and a strongly coupled CFT \cite{Maldacena:1997re}. Another example is the near horizon of the extremal black holes. In this case,  the enhanced symmetry in the near horizon geometry provides a description for extremal black hole in the context of Kerr/CFT \cite{Guica:2008mu,Hartman:2008pb} or entropy function \eg \cite{Sen:2008vm}.

In addition to symmetry enhancement, the near horizon of  generic extremal black holes with smooth horizon enjoys more interesting features: Firstly, they can  be considered as  a new class of solutions to the same gravity theory of the original black hole. These solutions usually  have a 2d maximally symmetric subspace.
As is proved in \cite{arXiv:0705.4214,arXiv:0806.2051,arXiv:0909.3462}, this 2d subspace is limited to be AdS$_2$ or 2d flat by imposing strong energy condition (see \cite{arXiv:1306.2517} for a review.). Generically, in  most studied extremal examples, it turns out that this 2d subspace is AdS$_{2}$. 
Secondly, they show a 
thermodynamic-like behavior which is called  Near Horizon Extremal Geometry (NHEG) dynamics \cite{arXiv:1310.3727,arXiv:1407.1992}.

In above discussion, there is an implicit assumption: Horizon is smooth and the horizon area remains non-zero in the extremal limit. However, there is another class of extremal black holes for which horizon area vanishes in the extremal limit such that
\be
A, \kappa \to 0, \quad \quad \frac{A}{\kappa}=finite,
\ee
where $A$ and $\kappa$ denote area and surface gravity of horizon, respectively. These kinds of black holes are called extremal vanishing horizon (EVH) black holes, vanishing of the horizon comes from the vanishing of one-cycle on the horizon \cite{arXiv:1107.5705} (for an incomplete list of EVH examples and their common features see \cite{hep-th/9905099,arXiv:0801.4457,arXiv:0805.0203,arXiv:0911.2898,arXiv:1010.4291,arXiv:1011.1897,arXiv:1112.4664,arXiv:1212.3742,arXiv:1212.4553,arXiv:1301.3387,arXiv:1301.4174,arXiv:1308.1478,arXiv:1407.7484,arXiv:1510.01209}).
One can study the near horizon geometry of EVH black holes similar to the generic extremal black holes. 

These near horizon EVH geometries are also solutions to the same original theory.
Interestingly enough, most EVH black holes admit a 3d subspace, which is generically pinching AdS$_3$,  as the near horizon geometry. In the case of near EVH black holes, this AdS$_{3}$ should be replaced with a BTZ black hole \cite{arXiv:1107.5705,arXiv:1504.03607,arXiv:1512.06186}. 
Thermodynamics of these BTZ black holes
represents the thermodynamics of original EVH black holes around the EVH limit \cite{arXiv:1107.5705,arXiv:1305.3157}. 

It is worthwhile to mention that the existence of the AdS$_3$ throat suggests a dual description in terms of AdS$_3$/CFT$_2$ for the near horizon physics of the EVH black hole and its excitations.
In fact in 4d case \cite{arXiv:1107.5705} it has been proven that the near horizon limit of a (near) EVH black hole is a decoupling limit. These suggest that a dual description for near horizon physics of EVH black holes in terms of  a 2d  CFT which is called EVH/CFT \cite{arXiv:1107.5705}.

The near horizon structure of EVH black holes has been studied in \cite{arXiv:1504.03607,arXiv:1512.06186}. A summary of these results is in the following theorems:

\begin{itemize}
	\item \textbf{ Theorem I.} \textit{The near horizon of any EVH black hole in Einstein-Maxwell-Scalar-$\Lambda$ theory which has a finite energy momentum tensor at the horizon has a 3d maximally symmetric subspace.}
	\item \textbf{ Theorem II.} \textit{The strong energy condition\footnote{We note that here the energy condition is on the matter energy momentum tensor \emph{without} cosmological constant.} excludes the 3d de Sitter space ($dS_{3}$) in the near horizon of EVH black holes which are asymptotically flat or AdS. }
\end{itemize}

These theorems do not exclude the existence of dS$_3$ and 3d flat spacetime in the near horizon geometry once the spacetime is asymptotically dS. Indeed, exploring this possibility is one of the main motivation of this paper. In other words, we are mainly interested in the near horizon structure of asymptotically dS EVH solutions.

The presence of a subspace with positive curvature in the near horizon of asymptotically dS black holes is not new. For example, the near horizon of the dS-Schwarzschild black hole solution in the extremal limit (when the cosmological and black hole horizons coincide) is $dS_2 \times S^n$ denoted as Nariai solution \cite{Nariai1,Nariai2,Nariai3,Ginsparg:1982rs,gr-qc/9606052}.

This paper is organized as follow. In the first section, after a short review on multi-spinning Kerr black hole in the presence of a cosmological constant, we will study the (near) EVH limit and its near horizon limit in  odd dimensions. In particular we will show that the 3d part of near horizon geometry of EVH Kerr-dS can be either AdS$_{3}$, dS$_{3}$ or 3d flat. In the case of the near EVH limit, we also study the relation between the thermodynamics of the EVH and the 3d part of the near horizon. In section two, we will study EVH limit of the cosmological soliton. In particular, we will show that the near horizon geometry enjoys a dS$_3$ subspace. The last section is devoted to discussions. 

\section{Kerr  black holes in higher dimensions}
In this section, we will apply the (near) EVH limit to the multi-spinning Kerr black hole in higher dimensions and in the presence of a (possible) cosmological constant \cite{hep-th/0404008,hep-th/0409155}.
This spacetime generalizes the single spinning 4d Kerr black hole to multi-spinning higher dimensional black hole and in the vanishing cosmological constant case, it reduces to the Myers-Perry solution \cite{Myers:1986un}. It solves the following Einstein equation
\bea\label{eom}
R_{\mu \nu}=(d-1)\lambda \; g_{\mu \nu}\;,
\eea
 in $d=2n+1+\alpha$ dimensions and its metric is given by\cite{hep-th/0404008,hep-th/0409155}
\bea\label{dS-MP}\label{bl}
ds^2 &=& - W\, (1 -\lambda\, r^2)\, dt^2
 + \frac{2\,m}{VF}\Bigl(W\,dt
 - \sum_{i=1}^n \frac{a_i}
  {\Xi_i}\, \mu_i^2\, d{\varphi}_i\Bigr)^2
 + \sum_{i=1}^n \frac{r^2 + a_i^2}{\Xi_i}\,\mu_i^2\,
    d{\varphi}_i^2 \nn\\
&&
 + \frac{VF\, dr^2}{V-2\,m}
 + \sum_{i=1}^{n+\alpha} \frac{r^2 + a_i^2}{\Xi_i}\, d\mu_i^2
 + \frac{\lambda}{W\, (1-\lambda r^2)}
    \Bigl( \sum_{i=1}^{n+\alpha} \frac{r^2 + a_i^2}{\Xi_i}
    \, \mu_i\, d\mu_i\Bigr)^2 \,.
\eea
Here, $\alpha$ is the ``even-ness" parameter such that it equals $1$ in even dimensions and 0 otherwise. In addition, one should set $a_{n+1}=0$ in even dimensions. In general, metric functions are
\be
\Xi_i\equiv 1+\lambda\, a_i^2\,,\qquad
W \equiv \sum_{i=1}^{n+\alpha} \frac{\mu_i^2}{\Xi_i}\,,\label{XiWdef}
\ee
\be
V\equiv r^{\alpha-2}\, (1-\lambda\, r^2)\, \prod_{i=1}^n (r^2 + a_i^2)\,,\qquad
F\equiv \frac{1}{1-\lambda\, r^2}\, \,
  \sum_{i=1}^{n+\alpha} \frac{r^2 \, \mu_i^2}{r^2+a_i^2}\,.\label{VFdef}
\ee
This solution is described by the parameters $m$ and $a_i$'s  which are respectively related to the mass and rotations. In the case of $m=0$, this solution is nothing but (A)dS. On the other hand, in the case of $a_i=0$ with non-vanishing $m$, spherical symmetry restores and this solution reduces to the known Schwarzschild-(A)dS black hole. This geometry is written in a coordinate system with ($n+\alpha$) number of latitudinal coordinates $\mu_i$  which are constrained by
\bea
\sum_i^{n+\alpha} \mu_i^2=1\;,
\eea
where $\mu_i \in [0,1]$ for $1\le i \le n$, $\mu_{n+1} \in [-1,1]$ for even dimensional cases
and $n$ number of azimuthal angular coordinates $\varphi_i \in [0,2\,\pi]$.
In general, this black hole solution may admit several horizons: inner and outer black hole horizons and a cosmological horizon for $\lambda>0$. All of them are determined by the roots of the following equation,
\bea\label{horizon}
V(r=r_h)=2\,m\;.
\eea
Let us consider a typical horizon $H$ which is specified by radius $r_{h}$. One may show this horizon is generated by the following Killing vector
\bea
\xi_{H}=\frac{\partial}{\partial t}+\Omega^i_H\frac{ \partial }{\partial \varphi^i}\; , \qquad \Omega^i_H=\frac{a_i (1-\lambda r_h^2)}{(r_h^2+a_i^2)}\;,
\eea
where $\Omega^i_H$ is the angular velocity of the horizon along $\varphi_i$. Then, the surface gravity computation gives
\bea\label{kappa}
\kappa_{H}=\frac{(1-\lambda\, r_{h}^2)}{4\,m}\,V'(r={r_h})\;.
\eea
Entropy and temperature of this horizon are
\begin{gather}\label{SandT}
S=\frac{A_{H}}{4 G_d} = \frac{ {\cal A}_{d-2}}{4 G_d}\; r_{h}^{\alpha-1}\,
  \prod_{i=1}^n
   \frac{r_{h}^2 + a_i^2}{\Xi_i}\,,\nonumber\\
T=\frac{\kappa_{H}}{2\pi}=\frac{1}{2\pi}\big{[}r_{h}(1-\lambda\, r_{h}^2)\,\big{(} \sum_{i=1}^n
\frac{1}{r_{h}^2 + a_i^2} +\frac{\alpha}{2r_h^2}\big{)} - \frac{1}{r_{h}}\big{]}\,,
\end{gather}
in which ${\cal A}_{n}$ is the volume of a unit n-sphere. The mass and angular momenta of this solution are given by \cite{hep-th/0408217}\footnote{Although the expressions of thermodynamic quantities  are given for negative $\lambda$ in that reference but those are valid for positive cosmological constant as well. This has been checked by calculating the charges via symplectic phase space method \cite{arXiv:1602.05575}.}
\begin{gather}\label{MandJ}
M=\frac{m\;\mathcal{A}_{d-2}}{4\,\pi\, G_{d}\, \prod_j \Xi_j}\big{(}\sum_{i=1}^{n}\frac{1}{\Xi_i}+\frac{\alpha-1}{2}\big{)}, \quad \quad
J_i=\frac{m\;\mathcal{A}_{d-2}}{4\,\pi\, G_{d}\,( \prod_j \Xi_j)}\frac{a_i}{\Xi_i}\,.
\end{gather}
Using these quantities, one can check the first law holds
\bea \label{FirstLaw}
\delta M=T\; \delta S+\sum_i \Omega^i_H \; \delta J_i\,,
\eea
here, $\delta$ denotes all possible variations in the parameter space of Kerr-(A)dS solution, \ie $\{m,a_1,a_2,\cdots , a_n\}$.
\subsection{(Near) EVH limit}
In the following, we explore the EVH limit of the Kerr black hole metric given by \eqref{bl}. From the equation \eqref{kappa}, it is clear that the extremal limit, $\kappa_H \to 0$, is simply given by the condition $V'(r=r_h)=0$ while $r_h$ is a root of \eqref{horizon}. To find the extremal vanishing horizon limit, we need to take vanishing horizon limit,  $A_{H} \to 0 $, as well, such that  $A_{H}/\kappa_{H}$ is fixed. Note that, generically, an extremal limit happens when two horizons degenerate and we define the vanishing horizon limit for the corresponding horizons.

In the case of solution \eqref{bl}, one can check that there is no such EVH limit in even dimensions. Therefore, in what follows, we only consider \emph{odd} dimensions and simply set $\alpha=0$.

To find the EVH limit, we simplify the entropy using the equation $V(r_h)=2\,m$ \eqref{horizon},
\bea
A_{H}=\frac{2\,m\, {\cal A}_{d-2} }{(1-\lambda r_h^2)}\left(\prod_{i=1}^n\frac{1}{\Xi_i}\right)\,r_h\;.
\eea
It is clear that entropy is not vanishing unless $r_h$ goes to zero. This can not be compatible with the equation which gives the location of the horizon \eqref{horizon}
\bea \label{VMhor}
\frac{(1-\lambda r_h^2)}{r_h^2}\prod_{i=1}^n (r_h^2+a_i^2)=2\,m\;,
\eea
 unless one of $a_{i}$'s is zero. We assume that the zero rotation parameter is along the $\varphi_1$ direction and set $a_1=0$. This assumption simplifies the surface gravity expression to
\bea
\kappa_{H} =\frac{(1-\lambda r_h^2)}{4\,m}\,V'(r_h)\Big|_{a_1=0}=\left(\sum_{i=2}^n \frac{1-\lambda r_h^2}{r_h^2+a_i^2}-{\lambda}\right)\, r_h\;,
\eea
which is also proportional to $r_h$. So, by setting $a_{1}=0$ and $r_h=0$  one can obtain an extremal black hole with vanishing horizon area.  To show that the ratio of $A_{H}$ and $\kappa_{H}$ remains finite, we need to do more careful analysis. Actually, from the above argument and eq. \eqref{VMhor}, one may deduce it is necessary to set $a_{1}=0$ first and then $r_h=0$. In other words, we should consider the following scaling limit \footnote{In general, one can scale $a_{i}$ with $\epsilon^\beta$ as far as $\beta\ge 2$. However, the near horizon limit of the near-EVH  with $\beta=2$ includes also $\beta>2$ cases.}
 \bea \label{EVHscaing1}
r_{h}=\rh\, \epsilon\,, \ \ \ a_1=a_0\, \epsilon^2, \, \, \epsilon\rightarrow 0\,.
\eea
Now, we can obtain surface gravity $\kappa_{H}$ and area of the horizon $A_H$ via these scaling limits
\bea \label{KappaA}
\kappa_{H} =- \big{(}\frac{a_0^2}{\rh^4}+\lambda_{3}\big{)}\rh \epsilon\,, \; \quad
A_{H} ={\cal A}_{d-2} \big{(}\prod_{i=2}^n\frac{a_i^2}{ \Xi_i}\big{)}\rh  \epsilon\;,
\eea
in which, for convenience, $\lambda_3$ is defined by
\bea \label{lambda3}
\lambda_3\equiv \lambda-\sum_{i=2}^{n}\frac{1}{a_i^2}\,.
\eea
 It manifestly shows  $\frac{A_{H}}{\kappa_{H}}$ is finite in $\epsilon \rightarrow 0$ limit. One may note that the surface gravity may become negative for some values of parameters. However, as we will discuss later, it is not a serious issue and one can show either the negative temperature  is due to cosmological horizon (for $\lambda>0$) or violation of extremality bound.

 Note that EVH limit \eqref{EVHscaing1} should be also compatible with $V(r_h)=2\,m$. Therefore, we  need to fix the value of $m$ in an appropriate way. Consequently, we add the proper scaling of $m$ to \eqref{EVHscaing1} and take the following limit as the EVH limit of Kerr black hole \eqref{bl} in odd dimensions
\bea \label{EVHscaing}
r_{h}=\rh \epsilon,\, \ \ \ a_1=\a1 \epsilon^2, \, \, \ m=\frac{1}{2}\prod_{i=2}^{n}a_{i}^2+\tilde{ m}\, \epsilon^2, \ \ \ \ \ \, \, \, \epsilon\rightarrow 0,
\eea
where the parameter $\m$ is given by
\bea \label{mtilde}
\tilde{ m}=\frac{\rh^2}{2} \left(\frac{a_0^2}{\rh^4}-\lambda_3\right)\; \prod_{i=2}^{n}a_i^2.
\eea
For later use, we also apply  this limit to the angular velocity and the momentum along $\varphi_1$
\bea
\Omega^1=\frac{a_0}{\rho_0^2}+\mathcal{O}(\epsilon^2)\,, \qquad J_1=\frac{\mathcal{A}_{d-2}}{8 \pi G_{d} }\prod_{i=2}^{n}\frac{a_i^2}{\Xi_i^2}\; a_0\,\epsilon^2+\mathcal{O}(\epsilon^4)\,.
\eea
One may note the expansions of $J_{i}$ and $\Omega^{i}$ along the other directions ($i \neq 1$) start from the zeroth order of $\epsilon$.

There is an interesting interpretation for $\m$. Indeed, one can simply check that this term does \emph{not} contribute to any quantities in the EVH limit. However, for the near EVH case, this term becomes important and changes the mass of the black hole above the EVH. In  other words, we can interpret $\m$  as excitations above the EVH surface in the parameter space \cite{arXiv:1305.3157,arXiv:1301.3387}. Moreover, by eliminating $\lambda_{3}$  between $\tilde{ m}$ and $\kappa_{H}$ in \eqref{KappaA}, we find $\tilde{ m}$ depends on the temperature $T=\frac{\kappa_{H}}{2 \pi}$ and $\frac{\a1^2}{\rh^2}\sim \Omega_{H}^{1}J_{1}$ in this way
\be \label{delta_m}
\tilde{ m} \; \epsilon^2=\frac{8 \pi G_{d} \prod_{i=2}^{n}\Xi_i}{\mathcal{A}_{d-2}}\Big{(} \frac{1}{2} T S+ \Omega_{H}^{1}J_{1}\Big{)}+\mathcal{O}(\epsilon^4 ).
\ee
Curiously, the expression inside the parentheses is exactly the Smarr mass formula for a 3d BTZ black hole. It suggests that the excitations of EVH black holes may be governed by a 3d gravity \cite{arXiv:1107.5705,arXiv:1305.3157}.  However, one should note that this result is independent of the $\lambda_{3}$.  We will come back to this point later.
\subsection{Horizon structure in the EVH limit} \label{horizon-structure}
In the previous subsection, we considered the EVH limit for a typical horizon $H$. However, the metric \eqref{bl} admits various types of horizons: cosmological, inner or outer horizon. In what follows, we investigate under which conditions cosmological horizon (if exists) coincides with a black hole horizon or two black hole horizons degenerate in the EVH limit. As a part of these conditions, the sign of $\lambda_3$ would be constrained in each situation. 

Let us consider the horizon at $r=0$, in the EVH limit,
\bea\label{exact-EVH}
a_1=0\,,\qquad  2\,m=\prod_{i=2}^{n}a_i^2\,.
\eea
In general, to specify the type of the horizon we analyze $V(r)-2m$  whose roots determine the locations of the horizons. Applying the EVH conditions \eqref{exact-EVH} to this expression gives
\bea\label{V-2m}
V(r)-2\,m=(1-\lambda\,r^2)\,\Pi-2\,m\,,\quad \Pi\equiv \prod_{i=2}^{n}(r^2+a_i^2)\,,
\eea
that $\Pi$ is polynomial function of $r$ and can be rewritten as a summation,
\bea\label{Pi-expansion}
\Pi=\sum_{p=0}^{n-1} C_p\,\left(r^2\right)^p\,,
\eea
where the $C_p$ coefficients are defined by
\bea\label{Coeff}
C_p\equiv\sum_{i_1<\cdots<i_{(n-p-1)}=2}^{n}a_{i_1}^2\,a_{i_2}^2\cdots \,a_{i_{(n-p-1)}}^2\, ,\qquad C_{n-1}\equiv 1\,.
\eea
In particular, for $C_0$ and $C_1$, we have
\bea
&&C_0=\prod_{i=2}^{n}a_i^2\,, \qquad C_1=C_0\,\sum_{i=2}^{n}\frac{1}{a_i^2}\,.
\eea
Substituting \eqref{Pi-expansion} into \eqref{V-2m}, we obtain
\bea
V(r)-2\,m
=(C_0-2\,m)+\sum_{p=1}^{n-1}\left(C_p-\lambda\, C_{p-1}\right)\,r^{2p}-\lambda\,C_{n-1}\,r^{2n}.
\eea

Again, using EVH conditions \eqref{exact-EVH}, we have $C_0=2\;m$ and by defining $C_n=0$, we arrive at the following form for the \emph{exact} expansion of $V(r)-2\,m$,
\bea\label{coeff}
V(r)-2\,m&=&\sum_{p=1}^{n}c_p\,r^{2p}
\,;\qquad c_p\equiv \left(C_p-\lambda\,C_{p-1}\right)\,.
\eea
Near the origin, $r=0$, the most dominant term of $V(r)-2\,m$ comes from the smallest power of $r$, \ie $r^2$, which is
\bea
c_1=C_1-\lambda\,C_0=C_0\,\left(\sum_{i=2}^{n}\frac{1}{a_i^2}-\lambda\right)=-\,2\,m\,\lambda_3\,,
\eea
and far from the origin, the term $r^{2n}$ is dominant with the coefficient $c_n=-\,\lambda$. Depending on the sign of $\lambda$, these coefficients can be positive or negative. In the following, we explore each case of  positive/negative $\lambda$ separately.

\subsubsection{Kerr-dS ($\lambda>0$)}
In the presence of positive cosmological constant, the Einstein equation admits spacetimes with a cosmological horizon. So, the type of multi-spinning Kerr-dS EVH  black hole horizons are  more complicated. There are three types of degenerate horizon: (i)  the outer horizon of black hole coincides with the cosmological horizon, (ii)  the outer horizon comes to the inner horizon and (iii) cosmological, outer and inner horizons coincide. Following the nomenclature of \cite{gr-qc/9806056} for the extremal 4d Kerr-dS, we also call them the \emph{Nariai}, \emph{cold} and \emph{ultra-cold} limit, respectively.

\begin{figure}[t]
	\centering
	\includegraphics[width=0.6\linewidth]{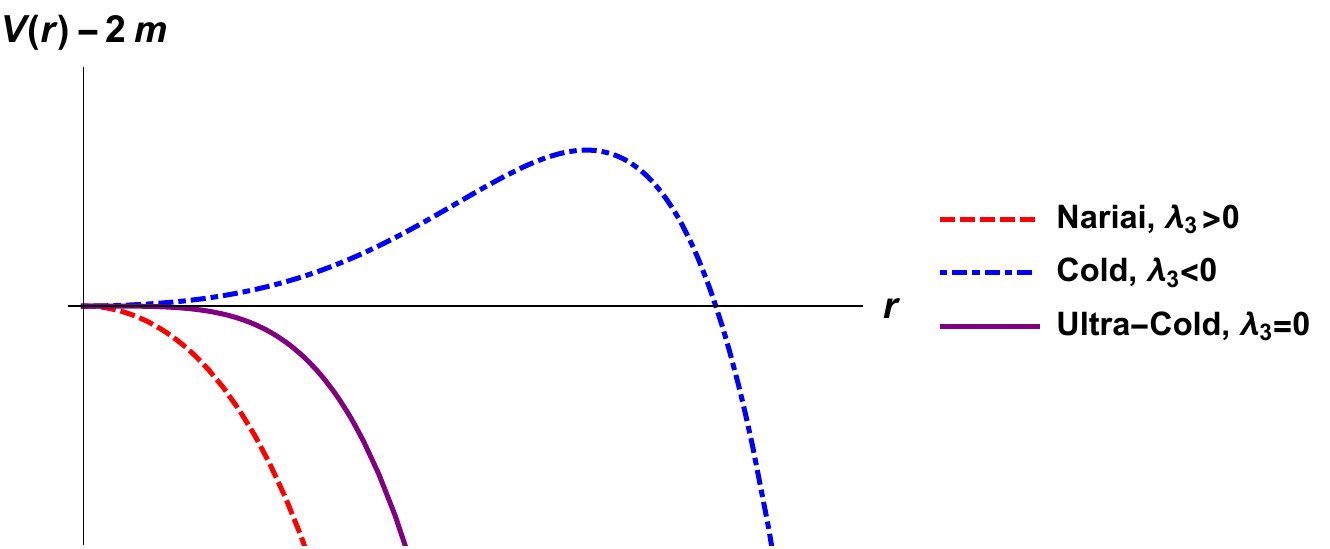}
	\caption{Roots of Kerr-dS black hole ($\lambda>0$) in the EVH limit.}
	\label{fig:pos-lambda}
\end{figure}

 \textbf{Nariai limit} For the asymptotically dS spaces, $c_n$ is negative, then for the ranges of parameters $a_i$'s where all $c_p$'s are negative, $V(r)-2\,m$ has no positive root except at $r=0$ (see Fig. \ref{fig:pos-lambda}). 
This condition,
\bea\label{Nariai-condition}
c_p<0\,\quad \forall\, p=1,2, \cdots, n\,,
\eea
also includes $c_1<0$ which translates to $\lambda_3>0$. As we will show in the next section it implies that the near horizon geometry have a dS$_3$ part. 

\textbf{Cold limit} This limit only happens when the spacetime has a cosmological horizon at $r>0$ (see Fig. \ref{fig:pos-lambda}).  As we will show in the following, it indicates $\lambda_{3}<0$. Descartes' rule of signs implies the existence of the cosmological horizon is only possible when at least a positive $c_{p}$ exists. \\For the case of 
\be \label{Cold-condition}
c_1>0\, \, (\lambda_3<0)\, ,\qquad c_p<0\,,\quad \forall p>1\,,
\ee
we have the possibility of at most a positive root for $V(r)-2\,m$. Accordingly, the slope and concavity of this function is also positive at the origin ($c_1>0$) and it goes to minus infinity in the large $r$ region ($c_n<0$), then it certainly has that positive root, at $r>0$. As we mentioned, this root is corresponding to the cosmological horizon. Therefore, in this case, the degenerate horizon at $r=0$ comes from the coincidence of black hole horizons (cold limit).

Now let us assume $c_{1}<0$ ($\lambda_3>0$) and that the cosmological horizon exists. So, $c_{p}$'s must be positive for some $p>1$. Using the asymptotic behavior of $V(r)-2m$  and its concavity at the origin and the cosmological horizon, one can deduce there is another root between the origin and the cosmological horizon. In this case, two inner black hole horizons are degenerate at $r=0$ (we do not study this case anymore since we are interested in the thermodynamics of outer black hole horizon).

\textbf{Ultra-cold limit(s)}  Up to here, we have assumed that none of the $c_p$'s are zero and the horizon at $r=0$ is just a double root or equivalently $r^2=0$ is a simple root (since $a_{i\neq 1}^2 > 0$). However, Kerr-dS admits $c_p=0$, then $r^2=0$ can be n-tuple root by requiring some exact relations between the parameters $a_i$ and $\lambda$.
The simplest case is $c_1=0$. In this case,
\bea\label{U-Cold}
c_1=0=\lambda_3\, \quad\Rightarrow\quad \lambda=\sum_{i=2}^{n}\frac{1}{a_i^2}\,.
\eea
Again, the condition,
\bea
c_p<0\,,\quad \forall \, p>2\,,
\eea
guarantees that the root at $r=0$, is the largest horizon (see Fig. \ref{fig:pos-lambda}). The only difference with the condition \eqref{Nariai-condition} is that $r^2=0$ is a double root now, instead of being a simple root. Thus, a cosmological horizon coincides with black hole inner and outer horizons which gives the ultra-cold limit.
We note that if $c_1\neq0$ and one of the other $c_p$'s vanishes, the root at $r^2=0$ does not change its type because the largest term in the origin comes from the smallest power of $r$. Therefore, to have higher q-tuple root at $r^2=0$, all $c_p$'s  for $p\leq q$ should be zero.

The summary of the results is
\bea\label{n-tuple}
&&\text{Double root at $r^2=0$}\;:\quad c_1=0\;,\nnr
&&\text{Triple root at $r^2=0$}\;~\,:\quad c_1=0\,\quad \text{and}\quad c_2=0\;,\nnr
&&\text{q-tuple root at $r^2=0$}\;:\quad c_p=0\,\quad \forall p<q\;.\nnr
\eea
\subsubsection{Kerr-(AdS) ($\lambda \leq 0$)}
\begin{figure}[t]
	\centering
	\includegraphics[width=0.6\linewidth]{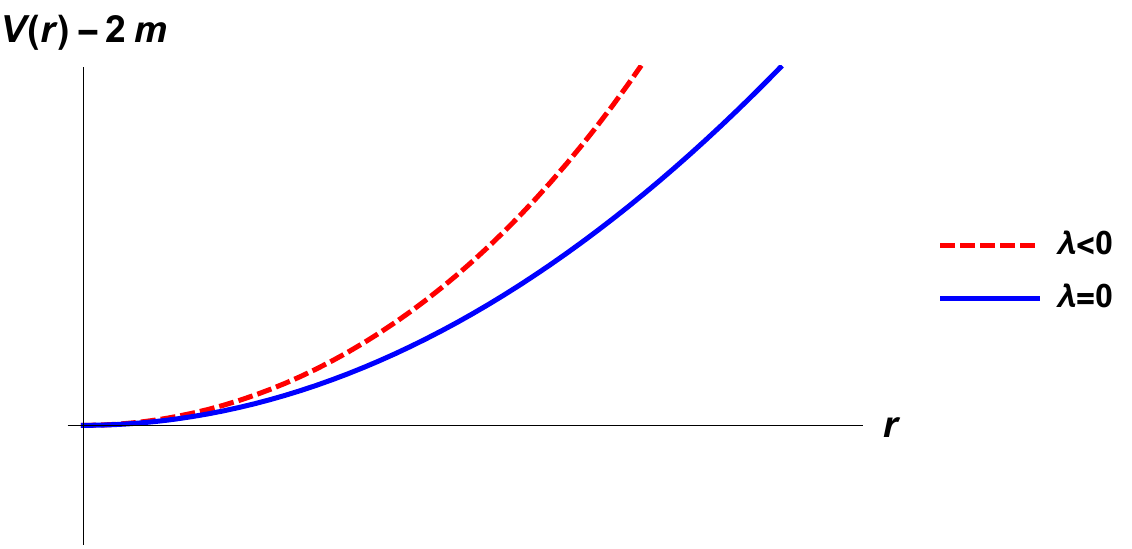}
	\caption{Root of Kerr black hole for $\lambda \leq 0$ in the EVH limit. There is no horizon for $r>0$. In these cases $\lambda_{3}<0$.}
	\label{fig:neg-zero-lambda}
\end{figure}

In these cases, $c_{p}$'s are always positive and there is no horizon for $r>0$ (see Fig. \ref{fig:neg-zero-lambda}). The degenerate horizon at origin is due to the coincidence black hole outer and inner horizons.
\subsection{Near horizon geometries}
The near horizon geometry of extremal 4d Kerr-(A)dS (non-vanishing horizon area) has been studied in \cite{arXiv:0811.4393,arXiv:0910.4587} and in the asymptotically AdS case in higher dimensions \cite{arXiv:0811.2225} \footnote{Recently, the near horizon of extremal Kerr-(A)dS-NUT solution in even dimensions has been studied in \cite{arXiv:1701.03950}. By setting NUT-charge to zero one can recover the extremal limit of even dimensions extremal Kerr.}.  For completion, we also mention the near horizon geometry of extremal Kerr for generic $\lambda$ in the appendix \ref{App-A}. In the following, we study the near horizon geometry of Kerr-dS black hole in the EVH limit.

It is more convenient to use the Kerr-Schild form of the metric \eqref{Kerr-Schild}. To obtain the near horizon limit of EVH, we apply the EVH limit \eqref{EVHscaing} and the following transformation 
\bea \label{nearhorizon}
r=\gamma\, \rho+r_{h},\, \ \  \bar{\tau}=\frac{v}{\gamma},\,
 \ \ \bar{\varphi}_1=\frac{\psi}{\gamma}, \, \ \  \bar{\varphi}_{i\geq 2}=\phi_{i}-\Omega_{H}^i\, \bar{\tau}, \, \  \ \  \gamma\rightarrow 0,
\eea
to the metric \ref{Kerr-Schild} and assume $\epsilon \ll \gamma$. In this case, the near horizon of EVH black hole reads as 
\begin{gather} \label{NH_EVH}
ds^2_{NH}=\mu_{1}^2\left[ \sigma\, \frac{\rho^2}{\l3^2}\, dv^2+2\,dv\,d\rho +\rho^2 \;d\psi^2\right]+h_{ij}({r_h})d\phi^i d\phi^j+k_{ij}({r_h})d\mu^i d\mu^j,
\end{gather}
where  $h_{ij}$ and $k_{ij}$ can be read from \eqref{dS-MP} or \eqref{Kerr-Schild} whose explicit forms are as follows,
\bea\label{metricH}
&&h_{ij}(r_h)=\frac{a_i^2\,\mu_i^2}{\Xi_i}\, \delta_{ij}+\frac{\mu_i^2\, \mu_j^2}{\mu_1^2} \frac{a_i\, a_j }{\Xi_i\,\Xi_j} \,,\nnr\nnr
&&k_{ij}(r_h)=\frac{a_i^2}{\Xi_i}\; \delta_{ij}+\,\lambda\; \frac{\mu_i\, \mu_j}{W}\, \frac{a_i^2\, a_j^2}{\Xi_i \Xi_j} \,,
\eea
where $i,j$ run from 2 to $n$. Also, $\mu_i$'s are restricted by $\sum_{i=2}^n\mu_i^2=1-\mu_1^2$. 
In addition, we introduced the 3d length scale $\l3$ for convenience and a sign bookkeeper $\sigma$ via
\be
\lambda_3=\frac{\sigma}{\l3^2}\,; \quad \sigma=(0,\pm 1)\,.
\ee
One may note that three coordinates $(v,\rho,\psi)$ make a 3d maximally symmetric spacetime and $\lambda_3$ determines its curvature. The 3d part is dS$_{3}$ when $\lambda_3> 0 $, 3d flat for $\lambda_3=0$ and AdS$_{3}$ once $\lambda_3 < 0 $. This 3d spacetime comes form joining 2d maximally symmetric subspace of near horizon of the extremal black hole and extra coordinate $\varphi_1$ due to vanishing horizon limit. 

Using the equation \eqref{lambda3}, it is clear that $\lambda_{3}$ is always negative or $\sigma=-1$, for $\lambda \leq 0$. This result is in agreement with theorems of \cite{arXiv:1504.03607,arXiv:1512.06186} which imply that for non-positive cosmological constant($\lambda\leq0$) EVH black hole in the Einstein-$\Lambda$ theory has an AdS$_3$ in the near horizon geometry. In this case, the near horizon geometry \eqref{NH_EVH} lies in the classified solutions with SO(2,2) symmetry in \cite{arXiv:1409.1635}. Besides, from equation \eqref{lambda3}, one can see the positivity of the cosmological constant $\lambda > 0$ for Kerr-dS admits 3d flat and dS$_3$ along with AdS$_3$. One may note that theorems of \cite{arXiv:1504.03607,arXiv:1512.06186} do not exclude these possibilities. Actually, one of the main motivations of this paper is to study this possibility.

Another interesting result is that $\lambda_3=0$ is exactly the condition \eqref{U-Cold} for $r^2=0$ to be a double root of $V(r)-2\,m=0$. Since all higher n-tuple roots assume vanishing of the second derivative, the near horizon EVH geometries for all of those cases include 3d flat spacetime.

We emphasize that this geometry is regular everywhere except at $\mu_1=0$ on which the Kretschmann scalar blows up. This is a typical features of EVHs \cite{arXiv:1107.5705,hep-th/9905099}
\subsubsection*{Near horizon of near EVH case}
In the previous discussion, we consider the near horizon limit for the case $\epsilon\ll\gamma$. However, in the $\gamma \sim \epsilon$ limit, we find a more general near horizon geometry which is called near horizon near EVH geometries. So, once again, we apply \eqref{EVHscaing} and \eqref{nearhorizon}, but this time we assume $\epsilon \sim \gamma$. Up to  a shift $ \rho\to \rho- \rh$, it leads to
\begin{gather}\label{NHNEVH1}
ds^2_{NH}=\mu_{1}^2\,ds_3^2+h_{ij}({r_h})d\phi^i d\phi^j+k_{ij}({r_h})d\mu^i d\mu^j,
\end{gather}
where $ds_3^2$ is defined by
\bea\label{ds3}
ds_3^2=-\;f(\rho)\;dv^2+2\;d\rho\,dv+\rho^2(d\psi-\frac{a_0}{\rho^2}\;dv)^2\,,
\eea
and $f(\rho)$ is given by
\bea
f(\rho)=\frac{(\rho^2-\rh^2)(-\sigma\,\rho^2-\;\frac{a_0^2\;\l3^2}{ \rh^2})}{\rho^2 \l3^2}\;.
\eea
This metric reduces to the near horizon metric of EVH \eqref{NH_EVH} for $\a1 = \rh =0$. The corresponding 3d part of the above geometry, depending on $\sigma$, is BTZ or Kerr-dS$_3$  \cite{hep-th/9806119} or 3d  flat space cosmology (FSC) \cite{hep-th/0203031,hep-th/0310099}. Metric functions $f(\rho)$ suggests all of these 3d spacetimes have horizon and one can attribute temperature or entropy to their horizon.\footnote{We note that using the transformation $dv=dt+\frac{d\rho}{f(\rho)}$, the metric takes the standard form of BTZ, Kerr-dS$_3$ and flat space cosmology metric,
\bea
ds^2=-f(\rho)\,dt^2+\frac{d\rho^2}{f(\rho)}+\rho^2\,(d\tilde{\psi}-\Omega\,dt)^2\,.
\eea } In what follows, we survey the thermodynamics of this 3d part. 

\subsection{Thermodynamics of the EVH near horizon }
As we mentioned before, the near horizon of EVH black hole admits a 3d maximally symmetric subspace which is replaced by a more general 3d spacetime in the near EVH. We take this 3d subspace as a solution to a 3d gravity which is obtained by a Kaluza-Klein reduction over the $\mathcal{M}_{d-3}$ manifold via reduction ansatz \eqref{NHNEVH1}. It is easy to show that the 3d Newton constant, $G_3$, is given in terms of d-dimensional Newton constant, $G_{d}$, as
\be\label{GDtoG3}
G_{3}=\frac{2\,\pi \,G_{d}}{\mathcal{A}_{d-2}}\prod_{i=2}^{n}\frac{\Xi_i}{a_i^2}\,.
\ee
Since the 3d metric has horizon, we can study its thermodynamic in the context of the mentioned 3d gravity. Using the standard methods, charges and chemical potentials of this 3d spacetime are obtained as 
\begin{gather}
M_{3}=\frac{a_0^2 \l3^2-\sigma\rh^4}{8 G_{3} \l3^{2} \rh^2} \;\epsilon\;,  \ \ \ \ S_{3}=\frac{\pi \rh}{2 G_{3}}\; \epsilon\;,  \ \ \ \ J_{3}= \frac{a_0}{4 G_{3}} \; \epsilon\;, \nn \\
T_{3}=\frac{1}{2 \pi \l3^2}\frac{-\sigma\rh^4-a_0^2 \l3^2}{\rh^3}\;,\ \ \ \ \Omega_{3}=\frac{a_0}{\rh^2}\;,
\end{gather}
where $\epsilon$ factor comes from periodicity of $\psi \in [0,2\pi \epsilon]$ due to the near horizon limit.
Interestingly enough, there is one to one correspondence between these quantities and the thermodynamic quantities of original black hole
\bea
&&T=\epsilon \,T_{3} \;, \ \ \ \ \Omega_{3}=\Omega^{1}_{H}\;, \nn \\
&&S=S_{3}\; , \ \ \ \ \ J_{1}=\epsilon\, J_{3}\;.
\eea
In addition, the mass of 3d spacetime is related to the excitation of mass parameter $m$ above the EVH limit, i.e $\tilde{m}$ given in \eqref{mtilde} through
\be
M_{3}=\frac{\mathcal{A}_{d-2}}{8\pi\,G_d\,\prod_{i=2}^{n}\Xi_i}\,\tilde{m}\, \epsilon\,.
\ee
As we mentioned below equation \eqref{delta_m}, it represents Smarr mass formula $M_{3}=\frac{1}{2}T_{3}S_{3}+\Omega_{3}J_3$. It is straightforward to see these quantities satisfy the first law of thermodynamics for 3d solution
\bea
\delta_{\perp} M_3=T_3\; \delta_{\perp} S_3+\sum_i \Omega_3 \; \delta_{\perp} J_3\,,
\eea
where $\delta_{\perp}$ refers to variations in the parameter space of 3d solution, \ie $\{a_0,\rho_0\}$. This is a sub-class of parametric variations of Kerr-dS which controls the distance from the EVH surface in the black hole parameter space, whereas, varying $a_{i\neq1}$ would maintain the solution EVH, in the EVH limit \eqref{EVHscaing}. In this sense, following \cite{arXiv:1301.3387,arXiv:1305.3157}, we call the first subclass \emph{normal} variations $\delta_{\perp}$ and the latter \emph{parallel} variations $\delta_{\|}$. In other words, it means the generic variation $\delta$ has a decomposition as $\delta=\delta_{\perp}+\delta_{\|}$. Using these, it is easy to see that $T \delta_{\perp}S=T_3 \delta_{\perp}S_3$ which implies the following relation between the first law of near EVH black hole and the corresponding first law for 3d space in the near horizon,
\bea
T\; \delta_{\perp} S\ &=&\delta_{\perp} M-\sum_i \Omega^i_H \; \delta_{\perp} J_i\,,\\ \nn
&\Downarrow&\nn \\ \nnr
T_3\; \delta_{\perp} S_3 &=&\delta_{\perp} M_3- \Omega_3 \; \delta_{\perp} J_3\;.
\eea
Indeed, it generalizes the relation between near EVH black holes thermodynamics and BTZ thermodynamics \cite{arXiv:1301.3387,arXiv:1305.3157} to general 3d locally maximally symmetric spaces.

One may note the temperature of the 3d spacetime is always negative  for $\sigma\geq 0$ ($\lambda_{3} \geq0$). From the 3d point of view, these cases present cosmological horizon of a Kerr-dS$_{3}$ and cosmological flat solution. For these spacetimes, the negativity of the cosmological horizon temperature is known \cite{hep-th/0407255}. Besides, from the original EVH black hole perspective, as we mentioned before, $\lambda_{3} \geq0$ is corresponding to Nariai and ultra-cold limits where in the both of them the cosmological horizon is involved. Therefore, one can relate the negativity of the temperature to cosmological nature of the horizon. In addition, for $\sigma=-1$ where the 3d part is a BTZ black hole, one must assume $\rh^2>\a1 \l3$ to preserve the extremality bound of the near EVH black hole and the corresponding BTZ black hole. Clearly, it implies the positivity of temperature.    

\subsection*{Cardy-like formula and black hole entropy}
We have seen the correspondence between thermodynamic quantities of the near horizon geometry and the EVH black hole.
For $\lambda_{3}<0$, the existence of AdS$_3$ geometry suggests a CFT$_{2}$ dual description for the physics of the near horizon geometry. In particular, one can apply Cardy formula to obtain the entropy of the near horizon BTZ black hole \cite{hep-th/9712251} which equals to the original near EVH black hole entropy\cite{arXiv:1107.5705}. In appendix \ref{cardy}, we study this case.

For $\lambda_{3}\ge 0$, where the 3d part of the near horizon is either locally flat or dS$_3$, there are several proposals for dual description of these geometries in the context of dS/CFT \cite{hep-th/0106247,hep-th/0110108,hep-th/0112218} and flat space holography\cite{arXiv:1208.4371,arXiv:1208.4372}. Although these proposal generically have some problems with unitarity, but one may apply their procedures and use the Cardy-like formula to Kerr-dS$_3$ or 3d FSC and obtain the entropy.
\section{Cosmological soliton}
In this section, we will study the EVH limit of cosmological soliton. This geometry is asymptotically dS and as we will show, and it also admits a dS$_3$ in the near horizon region of its EVH limit. The metric of the cosmological soliton in odd $d(=2n+1)$-dimensions is given by \cite{hep-th/0508200, hep-th/0508109}
\bea
ds^{2} = -g(r)\;dt^{2}+\frac{dr^{2}}{g(r)f(r)}+\left( \frac{r}{n}\right) ^{2}f(r)\Big( d\psi
+\sum_{i=1}^{n-1}\cos \theta _{i}\,d\phi _{i}\Big) ^{2}
+\frac{r^{2}}{2n}\sum_{i=1}^{n-1}d\Sigma_i^2,
\eea
where 
\begin{gather}
g(r)=1 - \frac{r^{2}}{\ell ^{2}}\;,\quad f(r)=1-\frac{a^{2n}}{r^{2n}},\; \quad
d\Sigma_i^2 = d\theta_i^2+\sin^{2}\theta_i \;d\phi_i^2,
\end{gather}
$\theta_i$ and $\phi_i$ parametrize ($n-1$) numbers of 2-spheres, so $\theta_i\in[0,\pi]$ and $\phi_i\in[0,2\pi]$.
This metric solves the Einstein equation \eqref{eom} with
\bea
\lambda=\frac{\sigma}{\ell^2}\,,\quad \sigma=0,\pm1\,,
\eea
where $\sigma=0,-1$ represent asymptotically flat and AdS spaces respectively which do not admit EVH limit. Thus, we only consider the $\sigma=+1$ case in the following.\\
This solution is described by two parameters $a$ and $\ell$. Since the factor $f(r)$ takes both negative and positive values and $\psi$ is periodic, one may worry about the existence of closed time-like curve (CTC). To avoid this issue, one needs to fix the range of parameters and coordinates. Besides, we are also interested in the static patch of the solution. Thus, to find a static CTC-free patch, we will analyze the metric in the following.  Let us consider two cases where $a>\ell$ or $a<\ell$ and determine the sign of each metric components. (We will come back to the case $a=\ell$ later when we study the EVH limit.) The summary of results is given in the following Tables \ref{table1} and \ref{table2} for $a<\ell$ and $a>\ell$ cases, respectively.
\begin{table}[h]
	\centering
	\begin{tabular}{l ccc}
		\hline\hline
		r&$0$\hspace{1.7cm}$a$ & \hspace{1.5cm}$\ell$&\hspace{1.2cm}$\infty$ \\ [0.5ex]
		\hline 
		$g(r)$ & $+$ &$+$ &$-$\\ [0.5ex]
		\hline 
		f(r)& $-$ &$+$ &$+$\\[0.5ex]
		\hline
		$|| \partial_{r}||^2$&$-$ &$+$ &$-$\\
		\hline
		$|| \partial_{t}||^2$& $-$ &$-$ &$+$\\
		\hline
		$|| \partial_{\psi}||^2$& $-$ &$+$ &$+$\\
		\hline\hline
	\end{tabular}
	\caption{\small{The $a<\ell$ case.}}
	\label{table1}
\end{table}

\begin{table}[h]
	\centering
	\begin{tabular}{l ccc}
		\hline\hline
		r&$0$\hspace{1.7cm}$\ell$ & \hspace{1.5cm}$a$&\hspace{1.2cm}$\infty$ \\ [0.5ex]
		\hline 
		$g(r)$ & $+$ &$-$ &$-$\\ [0.5ex]
		\hline 
		f(r)& $-$ &$-$ &$+$\\[0.5ex]
		\hline
		$|| \partial_{r}||^2$&$-$ &$+$ &$-$\\
		\hline
		$|| \partial_{t}||^2$& $-$ &$+$ &$+$\\
		\hline
		$|| \partial_{\psi}||^2$& $-$ &$-$ &$+$\\
		\hline\hline
	\end{tabular}
	\caption{\small{The $a> \ell$ case.}}
	\label{table2}
\end{table}
As is clear from Table \ref{table1}, in the case of $a<\ell$, this metric has time-like Killing vector ($\partial_t$) in the region $r\le\ell$ and so is static in that region. Meanwhile, it has CTC  along $\psi$ direction
in the region $r<a$, therefore we restrict our study to the region $a\le r\le\ell$ (it has also been studied in \cite{arXiv:1611.01131}). In this region, $r$ is a space-like coordinate and $t$ is time-like one. Evidently, $\psi$ and $\phi_{i}$'s are space-like coordinates in this region.

For the case of $\ell<a$, the region which does not include CTC is $r\ge a$. While the metric is static in the region $r\le \ell$ and does not have overlap with no-CTC region.
Therefore, we do not consider this case anymore.

The cosmological horizon of this solution is located at $r=\ell$, and the horizon topology is $S^{1}\times (S^2)^{n-1}$. 
One may note that when $\ell \to \infty$, this horizon disappears and the metric goes to the d-dimensional Eguchi-Hanson metric\cite{hep-th/0508200,hep-th/0508109}. 
\subsection{Thermodynamic quantities}
Using the symplectic phase space method\cite{arXiv:1602.05575,arXiv:1512.05584,arXiv:1606.04353}, the mass for this solution can be computed,
\bea\label{Mass}
M=\frac{k_n\; a^{2n}}{8\pi \,G_d \ell^2}\;; \quad k_n\equiv\frac{1}{2}\,\left(\frac{2\,\pi}{n}\right)^{n}\;,
\eea
where $G_d$ is the d-dimensional Newton's constant. The advantage of the symplectic phase space method is that it is not sensitive to the sign of cosmological constant and enables us to compute charges on co-dimension 2 surfaces at any radius, not necessarily at infinity. 
In addition, straightforward calculations for temperature and entropy reveal that
\bea\label{SvT}
T=-\, \frac{1}{2 \pi \ell}\;\sqrt{1-\left( \frac{a^2}{\ell^2}\right)^{n}}\;,\qquad S =\frac{ k_n\; \ell^{(2n-1)}}{2\;G_d}\; \sqrt{1-\left( \frac{a^2}{\ell^2}\right) ^{n}}\;.
\eea
Similar to Kerr-dS in the previous section, we assume the temperature of cosmological horizon is negative. This justifies the minus sign in above temperature. Given these thermodynamic quantities, it is easy to check the first law of thermodynamics, \footnote{In this paper, we fix $\lambda$ (cosmological constant) and do not consider the contribution of $\delta \lambda$ to thermodynamics. However, the authors of \cite{arXiv:1611.01131} use a different approach and take $\lambda$ as a variable.}
\bea
\delta M=T\;\delta S\;.
\eea
\subsection{Extremal vanishing horizon limit}
As is clear from the above, the relation between entropy and temperature of this solution  is given by $\frac{S}{T}=-\frac{\pi k_n \ell^{2n}}{G_d}$. Then, the EVH limit is simply obtained by $a \to \ell$, or more precisely
\bea\label{EVH-limit}
a=\ell \,(1-\frac{b^2}{2}\epsilon^2)\;,\quad \quad \epsilon \to 0\;.
\eea
The minus sign before $\epsilon^2$ ensures that we are taking $a \to \ell$ limit for $a < l$. It is worthwhile to mention, this solution does not admit an extremal limit with non-zero entropy, in contrast to what we usually expect in black hole physics. Near the EVH limit, temperature and entropy behave as
\bea\label{TS-nearEVH}
&&T=\tilde{T}\,\epsilon+\mathcal{O}(\epsilon^2)\,,\qquad \tilde{T}=-\frac{b\;\sqrt{n}}{2\,\pi\,\ell}\,,\nnr
&&S=\tilde{S}\,\epsilon+\mathcal{O}(\epsilon^2)\,, \qquad \tilde{S}=\frac{k_n\,b\,\sqrt{n}}{2\,G_d}\,\ell^{(2n-1)} \,.
\eea
From the above, it is clear that the ratio of temperature and entropy is finite in the $\epsilon\to 0$ limit.
In addition, the mass has the following expansion in the EVH limit,
\bea\label{Mass-nearEVH}
M=M^{(0)}+\, M^{(2)}\,\epsilon^2+\mathcal{O}(\epsilon^4)\,,
\eea
where
\bea
M^{(0)}=\frac{k_n\; \ell^{2(n-1)}}{8\,\pi \,G_d} \,,\qquad M^{(2)}=-\,n\,b^2\,M^{(0)}\,,
\eea
which remains non-zero ($M^{(0)}\neq0$) when $\epsilon \to 0$.  
Before closing this part, we would like to mention in this limit, vanishing of horizon area is a result of vanishing one-cycle on the horizon along the Killing direction $\partial_\psi$. It can be inferred by looking at the metric of the horizon in EVH limit
\bea
ds^2_{H}\sim \frac{\ell^2\; b^2}{n}\; \epsilon^2\left( d\psi+\sum_{i=1}^{n-1}\cos{(\theta_i)}\; d\phi_i  \right)^2+\frac{\ell^2}{2n}\;\sum_{i=1}^{n-1}d\Sigma_i^2+\mathcal{O}\left(\epsilon^2\right).
\eea
\subsection{Near horizon geometries}
In the EVH limit, $a \to \ell$, the width of static region goes to zero. For that narrow region, we study the near horizon geometry using 
\bea\label{nearhorizon-2}
r=a\,(1 +\frac{n}{2\, \ell^2}\,  \gamma ^2 \rho ^2)\,,\quad t=
\;\frac{\tau}{\gamma}\;,\quad \psi=\frac{\Psi}{\gamma }\;,\quad \gamma \to0\;,
\eea
along with the EVH limit \eqref{EVH-limit}. We note that this transformation is built such that we are taking the near horizon limit while we are still between $r=a$ and $r=\ell$.
This gives a geometry which includes two small parameters $\epsilon$ and $\gamma$. For $\epsilon \ll \gamma$, the resulting geometry is
\bea\label{NHEVH-Soli}
ds^2_{NH}=\left(\frac{\rho^{2}}{\l3^2}\;d\tau^2-\l3^2\,\frac {d\rho^{2}}{\rho^2}+\rho^2\; d\Psi^2\right)
+\frac{{\ell}^{2}}{2n}\sum_{i=1}^{n-1}d\Sigma_i^2\;,
\eea
where $\ell_3^2= \frac{\ell^2}{n}$. The expression in the parentheses describes a locally dS$_3$ spacetime whose radius is $\l3$. Using the transformation \eqref{nearhorizon-2} in the near EVH limit ($\epsilon \sim \gamma$), we get the near horizon near-EVH geometry
\bea\label{NHNEVH-Soli}
ds^2_{NH}=\left(f(\rho)\;d\tau^2-\,\frac {d\rho^{2}}{f(\rho)}+\;\rho^2\; d\Psi^2\right)+\frac{{\ell}^{2}}{2n}\sum_{i=1}^{n-1}d\Sigma_i^2\;,
\eea
where
\bea
f(\rho)=\frac{\rho^{2}}{\l3^2}-b^2\,.
\eea
One can compare this geometry with the generic 3d near horizon in \eqref{ds3}: they match by setting $a_0=0$, $\rho_0^2=\l3^2\,b^2$ and changing the coordinates as $dv=d\tau+\frac{d\rho}{f(\rho)},$ along with setting $\sigma=+1$ (since the geometry is locally dS$_3$).

We note that in both near horizon metrics \eqref{NHEVH-Soli} and \eqref{NHNEVH-Soli}, the range of coordinate $\Psi$ is $[0,2\,\pi\,\gamma]$. However, in contrast to what happens for the EVH limit of Kerr, near horizon metric for (near) EVH soliton is regular everywhere except at the origin (where the conical singularity of $\psi$ occurs).
\subsection{Thermodynamics of the EVH near horizon}%
Analogous to what we have done for the near horizon of EVH Kerr-dS in the previous section, we can define thermodynamic quantities for the 3d part of the near horizon metric. Therefore, we reduce the Einstein-Hilbert action on the ($d-3$) dimensional manifold to 3 dimensions via the metric ansatz \eqref{NHEVH-Soli}. After this reduction, the 3d Newton constant $G_3$ is obtained in terms of $G_d$ as
\bea
G_3=\frac{\pi \,G_d}{n\,k_n}\,\ell^{-2\,(n-1)}\,.
\eea
Then the mass, temperature and entropy of the remaining 3d space are
\bea
&& M_{3}=-\,\frac{b^2}{8\, G_3 }\,\epsilon\,,\qquad T_{3}=-\frac{b}{2\,\pi\,\ell_3}\,,\qquad S_{3}=\frac{\pi\,b\,\ell_3}{2\,G_3}\,\epsilon\,.
\eea
It is easy to check that the first law of thermodynamics holds for these quantities
\bea
\delta M_{3}=\,T_{3}\,\delta S_{3}\,.
\eea
Here, $\delta_{\perp}$ is the same as $\delta$, because the parameter space of 3d subspace in the near horizon and soliton are the same.
In other words, thermodynamics of this 3d subspace is induced by the cosmological soliton thermodynamics near the EVH limit, \eqref{Mass-nearEVH}, \eqref{TS-nearEVH}, with these scaling
\bea
M_{3}=\epsilon\, M^{(2)}\,,\qquad T_{3}=\tilde{T}\,,\qquad S_{3}=\epsilon\,\tilde{S}\,.
\eea
\section{Summary and discussion}
All the so far studied examples of EVH black holes 
\eg \cite{hep-th/9905099, arXiv:0801.4457, arXiv:0805.0203,arXiv:0911.2898,arXiv:1010.4291, arXiv:1011.1897, arXiv:1112.4664, arXiv:1212.3742, arXiv:1212.4553, arXiv:1301.4174, arXiv:1301.3387} 
and black rings \cite{arXiv:1308.1478,arXiv:1407.7484,arXiv:1510.01209} are asymptotically AdS or flat spacetimes and they have the AdS$_3$ factor 
in their near horizon geometries. Based on the theorems studied in \cite{arXiv:1504.03607,arXiv:1512.06186}, near horizon EVH geometries of asymptotically de Sitter spaces unlike anti-de Sitter spaces
can have either dS$_3$, 3d flat space or AdS$_3$ as a subspace.\footnote{In asymptotically AdS spacetimes, that subspace is restricted by the strong energy condition  to 
be only AdS$_3$ \cite{arXiv:1504.03607,arXiv:1512.06186}.}
Motivated by this, in this paper we analyzed the extremal vanishing horizon limit of two asymptotically de Sitter spacetimes.

In the first example, we studied the EVH limit of a d-dimensional Kerr-(A)dS black hole and see that, this limit can only occur in odd dimensions (similar to the EVH case of Myers-Perry black hole\cite{arXiv:1112.4664,arXiv:1308.1478}). 
Then we studied the near horizon behavior in this limit and observed that the near horizon EVH geometry enjoys a 3d maximally symmetric subspace. Depending on the sign of curvature of the 3d subspace, $\lambda_3=\lambda-\sum_{i=2}a_{i}^{-2}$, it can be dS$_3$, AdS$_3$ or flat. Thus, for asymptotically $AdS$ black holes ($\lambda<0$), the $\lambda_3$ is also negative and the 3d subspace can only be AdS$_3$. In the asymptotically flat case, this solution reduces to the known d-dimensional Myers-Perry black hole, so its near horizon contains AdS$_3$ ($\lambda_3<0$). However, in asymptotically de Sitter case ($\lambda>0$), $\lambda_3$ can be positive, zero or negative. For the highly spinning Kerr-dS black hole, $\lambda_3$ is positive and near horizon includes a dS$_3$ factor. 

In general, the Kerr-dS solution has at most $\left[\frac{d+1}{2}\right]$ roots for the horizon equation\eqref{horizon}. 
The largest one is the cosmological horizon. The EVH limit could be either the result of the degeneracy of a black hole outer horizon with its cosmological horizon (Nariai limit) or with its inner horizon (cold limit) or with both of them (ultra-cold limit). We discussed the necessary conditions for each of these limits in section \ref{horizon-structure}.

We also discussed the thermodynamics of the near EVH limit. Indeed, there are two types of fluctuations around the EVH black hole: those which remain in the EVH surface and those which make the solution parameters out of EVH surface (normal to the EVH surface in the parameter space). Allowing for the fluctuations normal to the EVH surface, we find near horizon near-EVH geometries. Depending on the sing of $\lambda_3$, we found Kerr-dS$_3$, BTZ or flat space cosmology factor in the near horizon near-EVH geometries. We also studied the thermodynamic behavior of these near horizon geometries. There is a one-to-one relation between these behaviors and the thermodynamics of the black hole itself. 
Since $\lambda$ is the parameter of the theory which is fixed and $a_i$'s are solution parameters which can be varied, it is interesting to study the phase transition between three possible 3d geometries via changing $a_{i}$'s in the context of original EVH black hole thermodynamics. 
On the other hand, we limited the variations to be \emph{normal} to the EVH surface and kept $\lambda_3$ and $G_3$ constant. It is possible to take $\lambda_3$ as a variable and study the thermodynamics of near horizon near-EVH geometries for generic variations which could be parallel or normal to the EVH surface. In this case one may study the thermodynamics of the near horizon geometry in the context of extended phase space  thermodynamics \eg \cite{arXiv:1608.06147,arXiv:1610.02038}.

 Presence of the AdS$_3$ factor in the near horizon geometry enables us to describe the physics of the black hole in the vicinity of its horizon by a 2d CFT in the context of AdS$_3$/CFT$_2$. In particular, the entropy of near horizon BTZ black hole is obtained via Cardy formula (see Appendix \ref{cardy}). 
When the 3d part of the near horizon metric is locally dS$_3$ or flat one may apply dS/CFT \cite{hep-th/0106247,hep-th/0110108,hep-th/0112218} and flat space holography \cite{arXiv:1208.4371,arXiv:1208.4372} proposals to study the near horizon geometry. Especially  for the proposed dS$_3$/CFT$_2$ in the Nariai limit, it is instructive to compare the result with what is obtained via Kerr/CFT in the Nariai limit of Kerr-dS \cite{Arxiv:0910.4587}\footnote{In  \cite{Arxiv:0910.4587} the authors only consider 4d Kerr-dS but its generalization would be straightforward.}.

In the second example, we considered another type of asymptotically de Sitter solution, the cosmological soliton. 
Using the symplectic phase space method\cite{arXiv:1602.05575}, we computed its thermodynamic quantities and integrable charges. 
Our result for the mass is independent of the radius of the co-dimension two surface on which the conserved charge is computed. Therefore, our result for the mass is different from the masses, $M_{in}$ and $M_{out}$ given in \cite{arXiv:1611.01131}.\footnote{We remind the reader that "in" and "out" refer to inside and outside the cosmological horizon in \cite{arXiv:1611.01131}.} The discrepancy can be a consequence of using the extended phase space thermodynamics that they have considered. 

An interesting observation is that the entropy of the cosmological soliton is proportional to its temperature. Thus, the extremal limit of this solution is already the EVH limit. In this sense, this is not a standard Nariai limit.

Unlike most other EVH black holes, the horizon of the cosmological soliton is smooth everywhere and free of curvature singularity. This is a counter example to the lore that EVH black holes are naked singularities.

In the near horizon of EVH cosmological soliton, we found only dS$_3$ subspace in contrast to the EVH Kerr-dS for which AdS$_3$ and 3d flat space is also possible. This dS$_3$ factor of the near horizon turns into Kerr-dS$_3$ in the near EVH limit. We also discussed thermodynamics of this type of geometries and its relation to soliton thermodynamics. After reduction on the ($d-3$) dimensional space, we get an asymptotically dS$_3$ space. Studying the dual CFT$_2$ (if it exists) is another interesting question which should be answered.

\section*{Acknowledgements}
We would like to thank M.M. Sheikh-Jabbari for fruitful comments and discussions. We also thank S.M. Nourbakhsh for her collaborations in the first stages of this project. We are thankful to H.R. Afshar and A. Aghajamali for their comments on the draft. The work of S.S. has been supported by the Allameh Tabatabaii Prize Grant of the National Elites Foundation of Iran. M.H.V. thanks the hospitality of ICTP, Trieste, where the final revisions of the paper were made.

\appendix
\section{Near horizon extremal geometry of Kerr-dS solution} \label{App-A}
 The near horizon of the extremal limit of more general Kerr-NUT-(A)ds spacetimes in even dimensions has been study recently in \cite{arXiv:1701.03950}. Here, we briefly note this limit for \eqref{bl}. Since the surface gravity of this black hole is proportional to $V'(r_h)$ \eqref{kappa}, the extremal limit simply is given by $V'(r_h)=0$. We are interested in the near horizon limit so we can expand the metric around $r_{h}$. In particular, the metric time-time component $g_{tt}$ is proportional to $(V(r)-2\,m)$ and it vanishes at $r_{h}$. Therefore, the near horizon expansion, $g_{tt}$ should start from $V''(r_h)$. The same argument also works for $g^{-1}_{rr}$.
 Changing the coordinates as
\bea\label{NH}
\varphi^i \to \phi^i=\varphi^i - \Omega^i_H\,t\;,\quad r-r_h=\gamma\, \rho\;,\quad t=\frac{\tau}{X \gamma}\;,\quad X=\frac{|V''(r_h)|}{4\,m}\, (1-\lambda \,r_h^2)\;
\eea
and taking the $\gamma \to 0$ limit, we find the near horizon extremal geometry,
\bea\label{NHEG}
ds^2=\frac{4\,m\, F}{V''}\left(-\rho^2 d\tau^2+\frac{d\rho^2}{\rho^2}\right)+h_{ij}({r_h})(d\phi^i-\rho\,k^i d\tau) \,(d\phi^j-\rho\,k^j\,d\tau)+k_{ij}({r_h})d\mu^id\mu^j\;,
\eea
where $h_{ij}$ and $k_{ij}$ can be read from \eqref{dS-MP} as
\bea \label{metricHext}
&&h_{ij}(r)=\frac{(r^2+a_i^2)}{(1+\lambda a_i^2)}\,\mu_i^2 \delta_{ij}+\frac{2\,m}{V\, F} \frac{a_i \mu_i^2}{(1+\lambda a_i^2)} \,\frac{a_j \mu_j^2}{(1+\lambda a_j^2)}\;,\nnr\nnr
&&k_{ij}(r)=\frac{(r^2+a_i^2)}{(1+\lambda a_i^2)}\, \delta_{ij}+\frac{\lambda}{W\,(1-\lambda r^2)} \frac{(r^2+a_i^2) \mu_i}{(1+\lambda a_i^2)} \,\frac{(r^2+a_j^2) \mu_j}{(1+\lambda a_j^2)}\,,
\eea
and $i,j$ run from 1 to $n$. In the near horizon geometry \eqref{NHEG}, $k^i$ is given by 
\bea
k^i=\frac{d\Omega^i}{dr}\Big|_{r=r_h}\,,
\eea
such that 
\bea
\Omega^i=\left(\frac{W}{1+g}\,\frac{2\,m}{V\,F}\right)\,\frac{a_i}{(r^2+a_i^2)}\,;\qquad g\equiv\frac{2\,m}{V\, F}\sum_{i=1}^{n}\frac{a_i^2\, \mu_i^2}{(1+\lambda a_i^2)\,(r^2+a_i^2)}\,.
\eea
(The horizon angular velocity $\Omega^i_H$, can be read from the above expressions for $\Omega^i$ on the horizon.)
Simple calculation shows that 
\bea
V''(r_h)=\frac{8\,m\, r_h^2}{(1-\lambda r_h^2)^2}\left(\frac{1-2\,\lambda\, r_h^2}{r_h^4}- \sum_{i=1}^n\frac{(1-\lambda r_h^2)^2}{(r_h^2+a_i^2)^2}\right)\,.
\eea

\section{Kerr-dS metric in the Kerr-Schild form}
The Kerr-Schild form of the Kerr-dS metric is given by\cite{hep-th/0404008}
\bea
ds^2=d\bar{s}^2+\frac{2m}{V\,F}(k_\mu dx^\mu)^2\,,
\eea
in which the de Sitter metric $d\bar{s}^2$ and the null one-form $k_\mu$ are as follows
\bea\label{Kerr-Schild}
d\bar{s}^2 &=& - W\, (1 -\lambda\, r^2)\, d\bar{\tau}^2
 + F\, dr^2
 + \sum_{i=1}^n \frac{\mu_i^2}{\Xi_i}\,(r^2 + a_i^2)\,
    d{\bar \varphi}_i^2 \nn\\
&&
 + \sum_{i=1}^{n+\alpha} (r^2 + a_i^2)\frac{d\mu_i^2}{\Xi_i}\,
 + \frac{\lambda}{W\, (1-\lambda r^2)}
    \left( \sum_{i=1}^{n+\alpha} (r^2 + a_i^2)\,\frac{\mu_i\, d\mu_i}{\Xi_i}
    \, \right)^2 \,,\nnr
k_\mu\,dx^\mu&=&W\,d\bar{\tau}+F\,dr-\sum_{i=1}^{n} \frac{a_i\,\mu_i^2}{\Xi_i}\,d\bar{\varphi}_i\,.
\eea
where, the functions $V$, $F$, $W$ and $\Xi$ are defined in equations \eqref{VFdef} and \eqref{XiWdef}.
Using the coordinate transformation\cite{hep-th/0404008}
\bea
&&d\bar{\tau}=dt+\frac{2\,m \, dr}{(1-\lambda\,r^2)\,(V-2\,m)}\,,\nnr
&&d{\bar \varphi}_i=  d{\varphi}_i+\frac{2\,m\, a_i\,dr}{(r^2+a_i^2)\,(V-2m)}\,,
\eea
one can get the metric \ref{dS-MP}.

\section{Black hole entropy from Cardy formula ($\lambda_3<0$)}\label{cardy}
For $\lambda_{3}<0$, there is a BTZ black hole in the near horizon of near EVH black hole. The standard AdS$_{3}$/CFT$_2$ shows Virasoro operators $L_0$ and $\bar{L}_{0}$ of CFT$_{2}$ are given in terms of mass $m_{BTZ}$ and angular momentum $J_{BTZ}$
\bea
L_0-\frac{c}{24}&=& \frac{1}{2}(\ell_3\,m_{BTZ}+J_{BTZ}),\nn \\
\bar{L}_0+\frac{c}{24}&=& \frac{1}{2}(\ell_3\,m_{BTZ}-J_{BTZ}),
\eea
where $c$ is the central charge of the corresponding Virasoro algebra. Following the seminal work of Brown-Henneaux \cite{Brown:1986nw} and taking into account the pinching periodicity of $\psi \in [0,2\pi \epsilon]$ one can show 
\be
c=\frac{3 \l3}{2 G_3}\epsilon=\frac{3 \l3 \mathcal{A}_{d-2}}{4\pi G_{d}}\prod_{i=2}^{n}\frac{a_i^2}{\Xi_i}\epsilon.
\ee
To keep $c$ finite, we should scale $G_{d}$ with $\epsilon$ \cite{arXiv:1011.1897}. Indeed, one can define the EVH limit via the following triple limits \cite{arXiv:1107.5705}

\be
A_{H},\; \kappa_{H},\; G_{d} \to 0, \quad \quad  \frac{A}{\kappa},\; \frac{A}{G_d} \;\; \text{finite},
\ee 

Now, we can obtain the finite entropy of the BTZ black hole via Cardy formula
\be
S=2\pi\sqrt{\frac{c}{6}\left(L_0-\frac{c}{24}\right)}+2\pi\sqrt{\frac{c}{6}\left(\bar{L}_0-\frac{c}{24}\right)}=\frac{\pi \rh}{2 G_{3}}=\frac{{\cal A}_{d-2}}{4 G_d} \big{(}\prod_{i=2}^n\frac{a_i^2}{ \Xi_i}\big{)}\rh,
\ee
which is exactly the entropy of near horizon BTZ and the corresponding near EVH black hole.

{}

\end{document}